\documentclass[prl,twocolumn,floatfix]{revtex4}
\usepackage{graphicx}

\newcommand{\bnvo}{$\beta-$Na$_{0.33}$V$_2$O$_5$}

\newcommand{\cm}{cm$^{-1}$}
\newcommand{\tna}{$T_{\mathrm {Na}}$}
\newcommand{\tmi}{$T_{\mathrm {MI}}$}
\newcommand{\figwid}{0.8\columnwidth}

\begin{document}
\title{Charge ordering signatures in the optical properties of \bnvo }
\author{C.~Presura, M.~Popinciuc, P.~H.~M.~van~Loosdrecht, D.~van~der~Marel,
        M.~Mostovoy,} %
\affiliation{Material Science Center, University of Groningen,
             Nijenborgh 4, 9747 AG Groningen, The Netherlands}
\author{T.~Yamauchi, Y.~Ueda}
\affiliation{Institute for Solid State Physics, University of Tokyo, Japan}
\date{\today}
\begin{abstract}
Temperature dependent optical spectra are reported for \bnvo. 
The sodium ordering transition at \tna = 240 K, and in particular 
the charge ordering transition at \tmi = 136 K strongly influence 
the optical spectra. The metal-insulator transition at \tmi\ leads to 
the opening of a psuedogap ($\hbar\omega = 1700$~\cm), and to the 
appearance of a large number of optical phonons. These observations, 
and the presence of a mid-infrared band (typical for low dimensional metals) 
strongly suggests that the charge carriers in \bnvo\ are small polarons.
\end{abstract}
\maketitle
 Low dimensional metals feature a variety of MI transitions
resulting from electron-phonon or electron-electron interactions. 
In charge density wave systems, like NbSe$_3$~\cite{challener} and
K$_{0.3}$MoO$_3$~\cite{travaglini,degiorgi91}, a MI transition is 
induced by a strong electron-phonon coupling (Peierls state). 
In Fe$_3$O$_4$ and Ti$_4$O$_7$~\cite{mott}, which are polaronic materials, 
a MI transition is induced by small polaron 
ordering (Verwey state). Finally, in systems lacking sufficiently strong  
electron-phonon interaction, such as for instance $(Me_2-DCNQI)_2Li_{1-x}Cu_x$~\cite{yamamoto,tajima}, 
a MI transition may occur due to charge ordering resulting from electronic coulomb 
interactions (Wigner crystal).
Materials of which the properties are dominated by electron-phonon interaction often 
show the appearance in the insulating state of a large number of phonons in the infrared spectrum along the chain direction. This phenomena, discussed by Rice~\cite{rice77} for the 1D organic peierls compound TEA(TCNQ)$_2$, has been found in several materials \cite{challener,travaglini,degiorgi91,degiorgi}, including those showing a Verwey transition . 
One of the intriguing features of all these materials is that they show a so called mid-infrared band in the optical spectra. It has been argued that for Fe$_3$O$_4$~\cite{degiorgi87} and many other materials~\cite{emin} the mid-infrared band can be understood as a polaronic response. However, also materials where Hubbard physics dominates may show a relatively strong mid-infrared band resulting from intra-band transitions\cite{yamamoto,tajima,fye,schultz}.  

The recent discovery\cite{yamada} of a clear metal-insulator
transition (MIT) in the vanadium bronze \bnvo\ has sparked a
revival of interest in this quasi one-dimensional (1D) metallic
system.
The room temperature crystal structure\cite{wadsley} of
$\beta-$Na$_{x}$V$_2$O$_5$ presents three 
crystallographically distinct Vanadium sites, labelled V$_1$, V$_2$ and
V$_3$. The Na atoms occupy lattice positions which can be
represented as a ladder along the $b$ axis (the chain direction).
For $x=1/3$ only 50 $\%$ of these lattice sites is
occupied, each rung hosting one Na atom randomly distributed
between the left and right hand side of the ladder. 
Each Na-atom donates one electron to the otherwise empty vanadium
$d$ bands. It is believed\cite{newyamada} that these electrons are
shared among the three V chains above the metal-insulator
transition $T_{\mathrm{MI}} = 136$~K , and that they condense on the 
$V_1$ zig-zag chain or the $V_2$ ladder below $T_{\mathrm{MI}}$. 
The metallic nature is rapidly lost for small deviations from $x = 
0.33$\cite{yamada}. The presence of metallic behavior for only a
sharply defined charge carrier concentration is different from
conventional MIT in $2$ or $3$ dimensions, where the metallic
phase occurs in a broad range of carrier densities above the
critical value. This unusual doping dependence probably results from 
the potential created by the neighboring Na atoms. 
Doping away from $x=1/3$ creates empty or fully occupied rungs on the 
Na ladder leading to large potential variations at the vanadium sites, and thus a 
decreased conductivity and eventually a localization of the charges.

\bnvo\ exhibits three phase transitions as a function of temperature: A sodium ordering transition at $T_{\mathrm{Na}}\cong 240$~K accompanied by a doubling of the unit cell along $b$, a MI transition at $T_{\mathrm{MI}}\cong 136$~K showing an additional tripling of the unit cell \cite{yamoura}, and finally an anti-ferromagnetic transition at $T_{\mathrm{CAF}}\cong 22$~K~\cite{yamada,vasiliev,ueda01}. 
The nature of the MI transition in \bnvo\ is presently unclear. There has been a suggestion that 
it results from {\em bi}-polaronic ordering \cite{kobayashi79,chakraverty}. 
But since the temperature dependent magnetic susceptibility \cite{yamada} shows that the spins of the charge carriers remain unpaired in the insulating state this is an unlikely scenario. Another suggestion is that the charge ordering transition is a Peierls transition \cite{obermeier}. This, however, is based on the assumption that \bnvo\ becomes a quarter filled system below $T_{\mathrm {MI}}$, which is contradicted by experiments showing a tripling of the unit cell along the $b$-direction. 
As is clear from the above, the origin of the MI transition as well as the nature 
of the charge carriers and the relevance of electron-phonon interactions remain open problems 
in \bnvo. This motivated the present study on the temperature dependent optical 
conductivity of \bnvo. 

 \begin{figure}[htb]
   \centerline{\includegraphics[width=\figwid,clip=true]{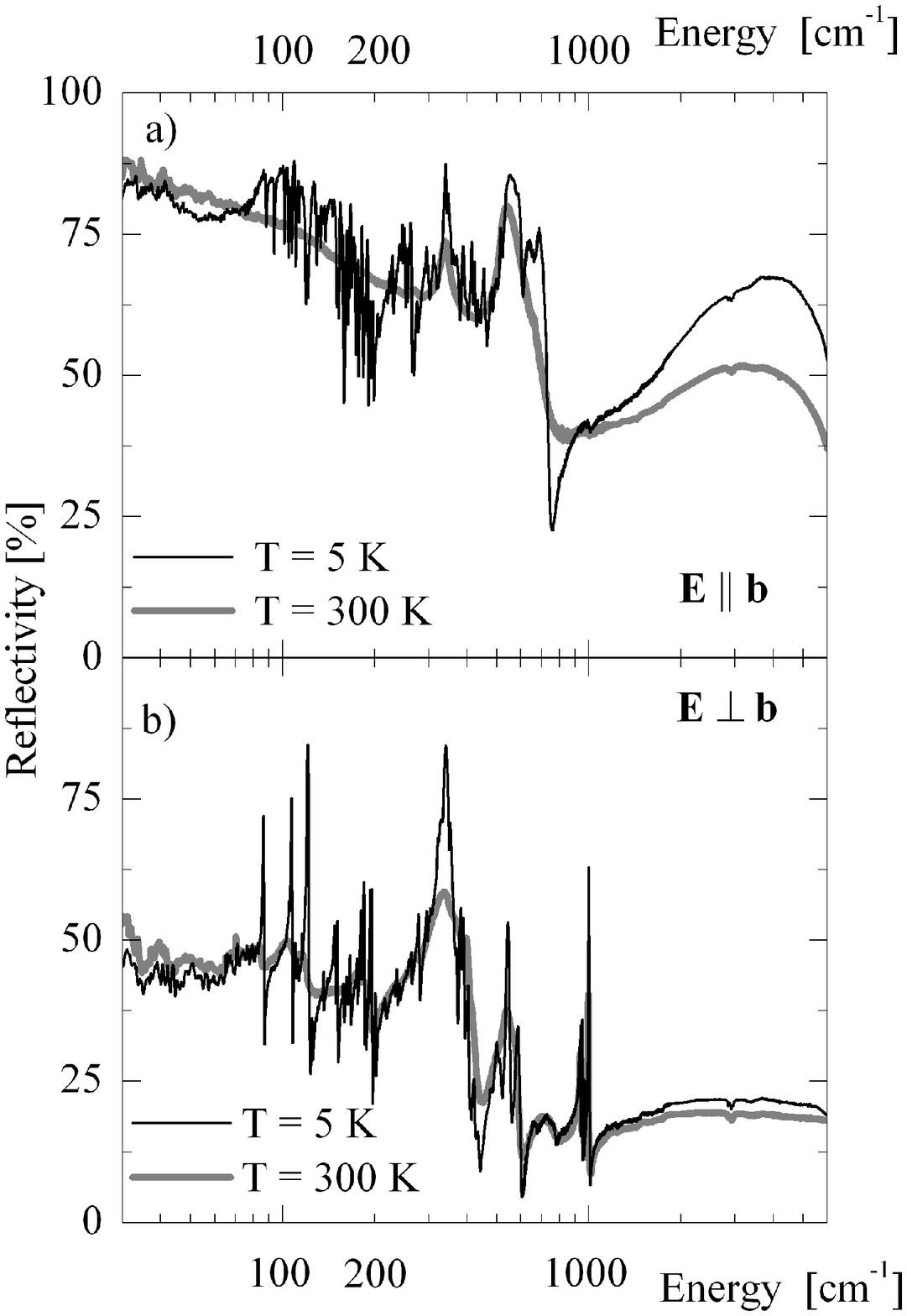}}
   \caption{
  Reflectivity spectra of \bnvo\ for $\mathbf{E}\| \mathbf{b}$ (a) 
  and $\mathbf{E}\perp \mathbf{b}$ (b).}
  \label{fig_reflectivity}
\end{figure}
\begin{figure}[htb]
     \centerline{\includegraphics[width=\figwid,clip=true]{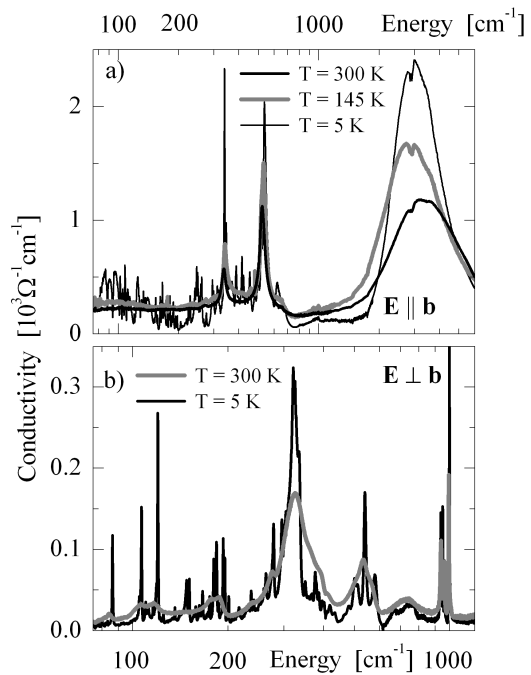}}
     \caption{
     The optical conductivity for $\mathbf{E}\|\mathbf{b}$ (a) and
     $\mathbf{E}\perp \mathbf{b}$ (b). The two small dips
     on the 3000~\cm peak are due to water absorptions. Note the different energy scales in     
     panel (a) and (b).}
  \label{fig_cond3000cm}
\end{figure}

Single crystals have been prepared as described in \cite{yamada}.
We measured the reflectivity in the range 20-6000 \cm\ as a
function of temperature with a polarizations both parallel and perpendicular 
to the $b$-direction ({\em i.e.} parallel to the conducting chains).
Figure~\ref{fig_reflectivity} shows the reflectivity spectra for
some selected temperatures. In addition, we used spectroscopic
ellipsometry to determine the dielectric function
from 6000 to 36000~\cm\ at room temperature. The optical conductivity was 
obtained by combining the reflectivity and ellipsometry data and performing 
a Kramers-Kronig analysis (see Fig. \ref{fig_cond3000cm}).
The one-dimensional nature of \bnvo\ is clearly reflected in the room temperature spectra. 
The $b$-direction shows a finite low frequency conductivity extrapolating to 200 $\Omega$\cm\ at zero frequency. This value is somewhat better than the DC value of 100 $\Omega$\cm\ measured previously~\cite{yamada}.
In contrast, the conductivity in the perpendicular direction shows a typical insulating behavior in that the conductivity extrapolates to zero for zero frequency. The other features observed in 
the spectra are several relatively sharp phonon lines, and more importantly a relatively strong 
mid-infrared band along the chain direction. 
The sodium ordering and in particular the MI transitions strongly influence the optical 
spectra. At the sodium ordering transition a small number of new phonon modes appear in the 
$E\perp b$ polarization. Their intensities have a slow temperature dependence reminiscent of an order/disorder transition and in good agreement with the temperature evolution of the satellites observed in X-ray diffraction experiments~\cite{kanai}. This is exemplified for a 
mode appearing at 990 \cm\ in figure 3b. More spectacularly, the MI transition 
results in the appearance of a large number of sharp phonon lines (more than 60 for the polarization along the $b$-direction, see figure 3a for a detailed view). 
This evidences the presence of strong electron-phonon coupling in \bnvo. Another indication for important electron-phonon interactions is the observation of strongly distorted phonon 
line shapes in the low frequency part ($<150$~\cm) of the conductivity along the $b$-direction (figure 2a).  The temperature dependence of the intensity of the phonons appearing in the insulating state exhibits a clear 2$^\mathrm{nd}$-order nature (see figure 3b, 950 \cm\ mode). 
A second change in the optical conductivity below the MI transition is the opening of a pseudo-gap below 1700 \cm\ in the $b$-direction (see figures 2a and 3c). This opening of the pseudogap, apparent as a clear decrease in the background intensity below \tmi,  is somewhat obscured by the appearance of the many phonon peaks in the insulating phase.

\begin{figure}[htb]
     \centerline{\includegraphics[angle=270, width=\figwid, clip=true]{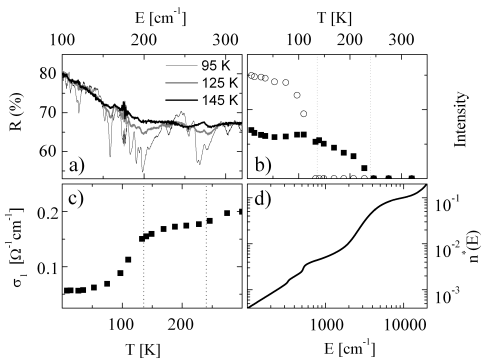}}
     \caption{a) detail of the reflectivity spectrum for $\mathbf{E}\|\mathbf{b}$\ for T=95, 125,                     and 145 K.
              b) Temperature dependence of the intensity of two phonons appearing in the 
                 $\mathbf{E}\perp \mathbf{b}$\ spectra at 990 \cm\ (full symbols) and 950 \cm\                       (empty symbols).
              c) Temperature dependence of the $\mathbf{E}\|\mathbf{b}$\ optical conductivity for                    $E=750$~\cm.
              d) Effective number of electrons calculated from the $\mathbf{E}\|\mathbf{b}$\       
                 optical conductivity using equation 1.
     }
  \label{fig3}
 \end{figure}
An important clue for the interpretation of the
optical spectra below 8000 cm$^{-1}$
is provided by the observation, that
while a gap opens in the optical spectrum and the
DC-resistivity below 136 K, the magnetic 
susceptibility is almost unaffected by the
metal-insulator transition.
If the gap were of the garden variety like in
silicon, the spin-fluctuations would become
strongly suppressed below the same temperature
where the optical conductivity becomes gapped.
However, in \bnvo\ there is no formation of a 
spin-gap when charge transport gets suppressed
below 136 K~\cite{yamada}. This aspect of the 
data reflects the presence of strong on-site
Hubbard-type repulsions between the charge
carriers, which in one-dimension causes the
electrons to behave like spinless fermions,
and it may thus be an experimental candidate for the
material which can reproduce the main theoretical expectations of
the partly occupied Hubbard chain: spin charge separation, and
fractionalization of the charge~\cite{zhang}. 
The relative independence of the spin and charge 
channels has been noticed before for the
Bechaard salts~\cite{vescoli}, the presence of a 
pseudogap in the optical excitations being opposed 
to the absence of a gap for spin excitations. 
Another important clue is provided by considering the
optical response of the charge carriers:
In a previous study\cite{kaplan}, a minimum in the $E||b$
reflectivity at 7200 cm$^{-1}$ has been attributed to a plasma
edge. We can see from Fig.~\ref{fig_cond3000cm}a, that the main
contribution to the oscillator strength associated with this
plasma minimum arises from the prominent mid-infrared band centered
at 3000 cm$^{-1}$. By integrating
the optical conductivity, and adopting the bare mass $m_e$\ 
for the electrons, we calculated the effective number 
of electrons displayed in the figure 3d) using
\begin{equation}
\label{sumrule}
 8 \int_0^{\omega} \sigma_1(\omega') d\omega' =n^{*}(\omega)
 \frac{4\pi n_V e^2}{m_e}.
\end{equation}
The integrated spectral weight of the mid-infrared feature 
corresponds to $n^{*}(10000 \mathrm{cm}^{-1}) \simeq 0.10$\ electrons per V atom, 
which is rather close to the nominal chemical
doping of $n=0.166$\ electrons per V atom. (The difference can
be easily understood from the fact that in these transition metal
oxides the effective mass of the electrons is about 2$m_e$.) 
The high oscillator strength of the
mid-infrared peak as shown by the value of $n^*$\ shows 
that it arises from the doped charge carriers.  
In contrast the spectral weight of the low
frequency part up to 1500~\cm\ is about 10$\%$\ of the total
mid-infrared feature (see inset Fig.\ref{fig_cond3000cm}c).
Finally, the relative intensity of the low frequency spectral weight
(in comparison to the 3000~\cm\ peak) is almost independent of temperature above \tmi.

From studies of the Hubbard model in one dimension we know, that
part of the intra-band spectral weight shows up as a band of
mid-infrared excitations. However, these studies have also demonstrated, that
for doping far away from half-filling of the Hubbard band the
intensity of the mid-infrared band is less than 20 \% of
Drude spectral weight\cite{fye,schultz}. This
rules out an interpretation of the mid-infrared peak in \bnvo\ in
terms of a pure Hubbard model. At the same time we underscore
the crucial role of Hubbard-type correlations for the 
independence of the spin response from the charge-gap in this 
material.

The most trivial explanation of the 3000\cm\ peak would be that 
it is a direct transition between bands
which are formed as a result of the Umklapp-potential of the
Na-superlattice below 240 K. However, the potential landscape
caused by the Na ions becomes randomly ordered above 240 K. Although
even a random potential would give rise to a mid-infrared peak, the
position of the mid-infrared peak would become strongly
temperature dependent in such a scenario, in contrast to
our experimental observations.

The remaining candidate for the mid-infrared band is to assume
that the charge carriers are small polarons.
Derived basically from the Frank-Condon model, the small polaron
peak~\cite{emin} can be viewed classically as an instantaneous
transition from a localized state to a neighboring localized state
in a rigid ionic environment. The environment responds to the new electronic configuration by emitting a wave package of multi-phonon oscillations, 
the envelope of which corresponds to the line shape of a polaron.  
It acquires not only a peak at several 
times the frequency of phonons, but also a finite conductivity at
low frequencies. 
The small polaron optical line shape, being influenced by the movement of ions, 
depends strongly on temperature~\cite{emin}. This is indeed what we observe 
for the mid-infrared feature in its high frequency part (see
Fig.~\ref{fig_cond3000cm}a). 

The low frequency part is unchanged
down to the $T_{\mathrm{MI}}=136K$, in opposition to what is
expected\cite{emin} and measured\cite{park}  for the small
polarons. This unexpected behavior may have a connection to the
disorder potential created by the Na atoms, which could smear out
the influence of the temperature at low frequencies. If so, it
would suggest that some disorder in the Na positions exists even below
\tna, in other words that the ordering of the Na atoms takes place
gradually. This is confirmed by the measured intensity of the 
990 cm$^{-1}$ phonon which gradually develops in the insulating 
direction  below \tna, being 
fully developed at \tmi\ (see figure 3b).  They
both suggest that the MIT takes place at a temperature where the
Na atoms are fully ordered. This strengthens the
argument that the Na potential influences strongly the movement of
the electrons on the V chains.

To summarize the interpretation of all optical features: Both the
mid-infrared peak at 3000 \cm, and the observation of strong
electron-phonon coupling support the picture that the charge 
carriers in \bnvo\ should be regarded as small polarons. The
strong Hubbard-type interactions are responsible for the
observed independent behavior of the spin- and charge
channels at the metal-insulator phase transition. Below the
phase transition the insulating state is a charge
ordered phase. Future experiments will have to establish the
detailed nature of the charge ordered state. Although
in a way a regular array of polarons also represents a
CDW, a distinguishing feature in this case, is that in an
ordinary CDW the spin- and charge sectors should be gapped
simultaneously, which clearly does not happen in \bnvo\ at 136 K, 
where the spin orders at a still lower temperature. The
nature of the crystallographic phase transition at 136 K ({\em i.e.}
the tripling of the unit cell along the chains) suggests, that
below 136 K the charges have become ordered with a commensuration
of order 3 on the three different types of V-chains and ladders in
the unit cell. This would imply that the doped charges are either distributed
equally over all V-atoms or in a 2/3-1/3 ratio over the V$_1$\ and V$_2$ atoms, 
resulting in a high degree of dilution. 
This a favorable condition for the formation of small 
polarons, consistent with the above interpretation of the spectra.

This work was supported by the Dutch Foundation for Fundamental Research on Matter (FOM) 
and by INTAS (99-155). 
We greatfully acknowledge A.T.~Filip for the polarized microscopy checks and G.~Maris and  T.~T.~M.~Palstra for fruitful discussions.

\end{document}